\documentclass[conference]{IEEEtran}
\IEEEoverridecommandlockouts
\usepackage{cite}
\usepackage{amsmath,amssymb,amsfonts}
\usepackage{algorithmic}
\usepackage{graphicx}
\usepackage{textcomp}
\usepackage{xcolor}
\usepackage{tikz}
\usepackage{bm}
\usepackage[]{algorithmic}
\usepackage{algorithm}
\usetikzlibrary{positioning}
\graphicspath{ {images/} }
\newtheorem{c1}{Claim}
\newtheorem{t1}{Theorem}
\newtheorem{l1}{Lemma}

\newtheorem{d1}{Definition}
\def\BibTeX{{\rm B\kern-.05em{\sc i\kern-.025em b}\kern-.08em
    T\kern-.1667em\lower.7ex\hbox{E}\kern-.125emX}}
    
\begin{document}

\title{An Optimal Vector Clock Algorithm for Multithreaded Systems\\
{\footnotesize }
\thanks{This work was partially supported by NSF CSR-1563544,CNS-1812349}
}

\author{\IEEEauthorblockN{1\textsuperscript{st} Xiong Zheng}
\IEEEauthorblockA{\textit{Electrical and Computer Engineering
} \\
\textit{The University of Texas}\\
Austin, USA \\
zhengxiongtym@utexas.edu}
\and
\IEEEauthorblockN{2\textsuperscript{nd} Vijay K. Garg}
\IEEEauthorblockA{\textit{Electrical and Computer Engineering} \\
\textit{The University of Texas}\\
Austin, USA \\
garg@ece.utexas.edu}
}

\maketitle

\begin{abstract}
Tracking causality (or happened-before relation) between events is useful for
many applications such as debugging and recovery from failures. Consider a
concurrent system with $n$ threads and $m$ objects. For such systems, either
a vector clock of size $n$ is used with one component per thread or a
vector clock of size $m$ is used with one component per object.
A natural question is whether one can use a vector clock of size strictly less than the minimum
of $m$ and $n$ to timestamp events. We give an algorithm in this paper that
uses a hybrid of thread and object components. Our algorithm is guaranteed to return the minimum number of components necessary for vector clocks. We first consider the case when the interaction between objects and threads is statically known. This interaction is modeled by a thread-object bipartite graph. 
Our algorithm is based on finding the maximum bipartite matching of such a graph and then applying K\"{o}nig-Egerv\'{a}ry
Theorem to compute the minimum vertex cover to determine the optimal number of components necessary for the vector clock. We also propose two mechanisms to compute such an vector clock when computation is revealed in an online fashion. Evaluation on different types of graphs indicates that our offline algorithm generates a size vector clock which is significantly less than the minimum of $m$ and $n$. These mechanisms are more effective when the underlying bipartite graph is not dense.

 \end{abstract}


\section{Introduction}

A fundamental problem in parallel and distributed systems is to determine the order
relationship between events of a distributed computation as defined by
Lamport's \emph{happened-before relation}
\cite{Lam:1978:CACM}. The problem arises in many areas including
debugging and visualization of parallel and distributed programs. 
%

Vector clocks, which were introduced independently by Fidge
\cite{Fid:1988:ACSC,Fid:1989:WPDD,Fid:1991:Com} and Mattern
\cite{Mat:1989:WDAG}, and their variants \cite{MarSab:1994:DC} are widely used to capture the causality between events in parallel and distributed systems. To capture the causality, each event is timestamped with the
current value of the local vector clock at the time the event is
generated. The order relationship between two events can then be
determined by comparing their timestamps. A vector clock contains one
component for every process in a distributed system. This results in message and
space overhead of $N$ integers in a distributed system consisting of
$N$ processes. In shared-memory based systems, there are two kinds of
vector clocks. Consider a
concurrent system with $n$ threads and $m$ objects. For such systems, either
vector clocks of size $n$ is used with one component per thread or a
vector clock of size $m$ is used with one component per object.
A natural question is whether one can use a vector of size strictly less than the minimum
of $m$ and $n$ to timestamp events. 
We show that this is indeed possible in this paper.

\begin{figure}[htbp] 
\centering
\includegraphics[width=0.5\textwidth]{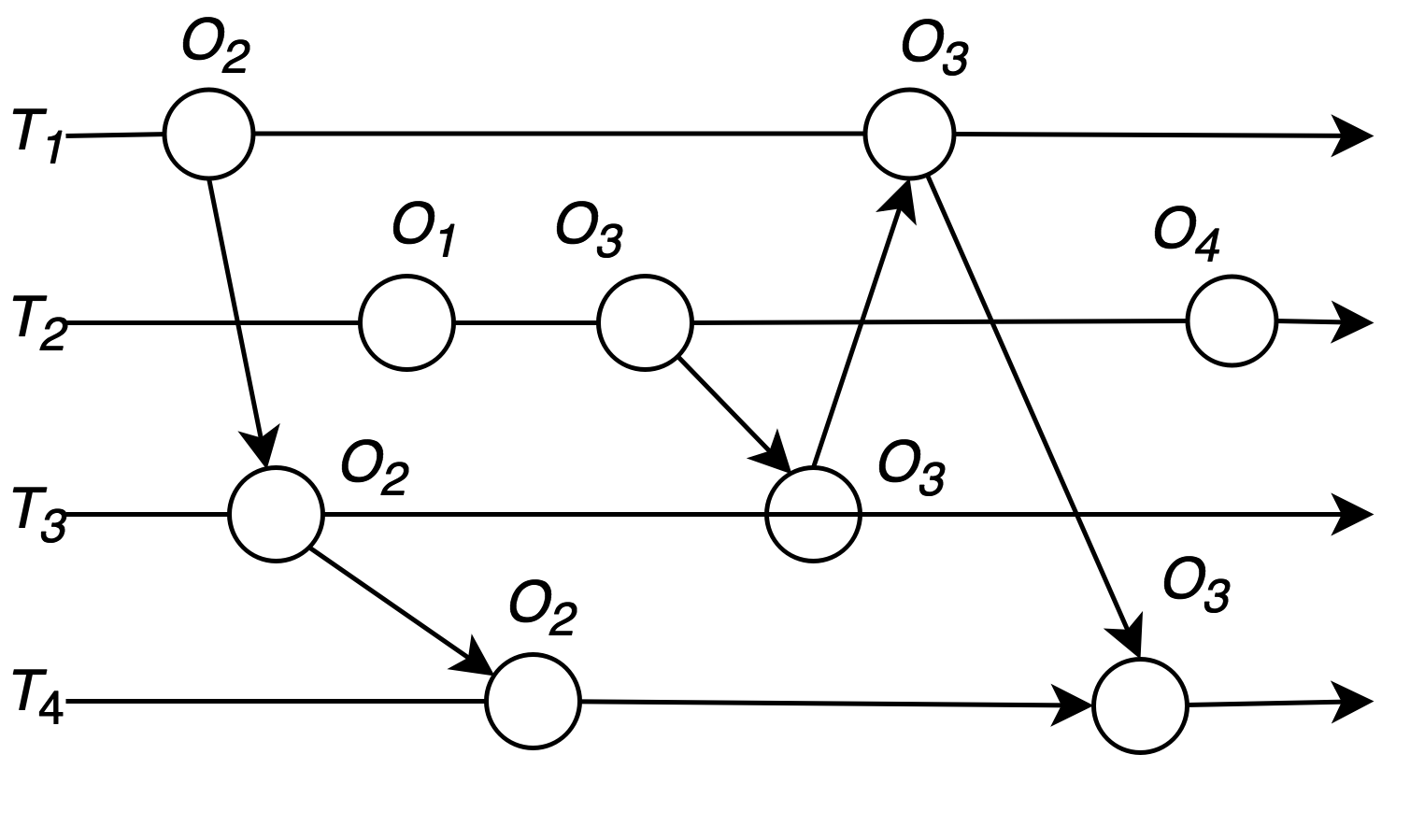}
\caption{A computation of threads operating on shared objects}
\label{fig:eg}
\end{figure}

Consider the example in Fig.~\ref{fig:eg} to understand why it is feasible to use a smaller vector clock to order events. In this example, $T_1, T_2, T_3, T_4$ are threads, $O_1, O_2, O_3, O_4$ are objects which are used by the above threads. Each circle represents an operation. To order those operations, traditionally, either all threads or all objects are exploited as components of vector clock so as to indicate the causal order. However, notice that all the operations are related to either thread $T_2$ or object $O_2$ or object $O_3$. Therefore, we can use a vector clock composed of $T_2$, $O_2$, and $O_3$ to timestamp all events. This mixed-vector-clock has the size of 3 which is smaller than the number of threads and the number of objects. This example answers the question that a vector of size strictly less than minimum of $m$ and $n$ can be obtained to timestamp a computation. 

A computation of threads operating on objects can be represented as a bipartite graph and the minimum vertex cover of a bipartite graph is no greater that the minimum of the number of left nodes and number of right nodes. Recall the definition of a vertex cover of a graph: a set of vertexes such that each edge in the graph is incident to at least one vertex of the set and notice that an edge in a bipartite graph actually corresponds to an event in a computation. Therefore, by converting such a computation to a thread-object bipartite graph (to be introduced in section III), we can make use of the minimum vertex cover of this bipartite graph to timestamp the events in the computation. 
Thus, the minimum vertex cover is used to determine the components of the vector clock for this computation. On this basis, we give an offline algorithm that makes use of a hybrid of thread and object components which is smaller in size than traditional vector clocks to timestamp a computation. Our offline algorithm is guaranteed to return the
minimum number of components necessary for a vector clock.
This algorithm is based on finding a maximum bipartite matching in the thread object bipartite graph and then applying K\"{o}nig-Egerv\'{a}ry
Theorem\cite{BondyMurty:1976:Book} to determine the optimal number of components necessary for the vector clock. 

We also consider the case when the interaction between threads and objects is not known {\em a priori}.
We propose two mechanisms to address this problem and compare performance with the traditional solution.


In summary, this paper makes the following contributions:
\begin{itemize}
\item
We introduce the notion of a mixed-vector-clock that satisfies the vector clock condition
with fewer than thread-based or object-based clock.
\item
We give an optimal algorithm to determine which threads and objects should be used for a mixed-vector-clock.
\item
We give two mechanisms to compute the mixed-vector-clock when events of a computation is coming in a online fashion. 
\end{itemize}

\section{System Model and Notation}
\label{sec:model}
In this section, we present our model of a concurrent system.
The system consists of $N$ sequential processes (or threads) denoted by $P = \{p_1,
p_2, \ldots , p_n\}$ performing operations on $m$ objects denoted by $Q = \{q_1, q_2, \ldots, q_m \}$. In the remaining section, we use threads or processes interchangeably. 
A \emph{computation} in the happened before model is defined
as a tuple $(E, \rightarrow)$ where $E$ is the set of events and $\rightarrow$
is a partial order on events in $E$. Each process executes a sequence of events.
Each event is performed on a single object. We assume that all operations on any single object
are performed sequentially (for example, by using locks).
 For
an event $e \in E$, $e.thread$ denotes the process on which $e$ occurred and
$e.object$ denotes the object on which $e$ occurred.

Then Lamport's happened-before relation ($\rightarrow$) on $E$ is the smallest transitive relation 
such that:

1. If $e.thread = f.thread$ and $e$ immediately precedes $f$ in the
sequence of events in process $e.thread$, then $e \rightarrow f$.

2. If $e.object = f.object$ and $e$ immediately precedes $f$ in the sequence of events on the object $e.q$, then
$e \rightarrow f$ \\


Two events $e$ and $f$ are said to be \emph{comparable} if $e \rightarrow f$ or
$f \rightarrow e$. If $e$ and $f$ are not comparable, they are said to be
\emph{concurrent} and this relationship is denoted by $e \parallel f$. 

We define process-object graph as the undirected bipartite graph $G=(P, Q, T)$ where
$T$ is the set of edges between the set of processes $P$ and the set of objects $Q$ defined as
\[ T = \{ (p,q) | \mbox{the object $q$ is accessible to process $p$} \} \]

In most applications, we do not expect $G$ to be dense, i.e, a process typically has references to only a small subset of objects.

%
The set of events $E$ with the order imposed by Lamport's happened before
relation defines a partially ordered set or \emph{poset}. A subset of elements $C
\subseteq E$ is said to form a \emph{chain} iff $\forall e,f \in C: e
\rightarrow f $ or $ f \rightarrow e$. By our definition of processes
all operations done by a single process form a chain. Similarly, all operations done
on a single object also form a chain.


Process-based vector clocks maintain a vector $V$ of size $|P|$ with each process 
and object. Whenever, a process $p$ executes an operation $e$ on object $q$ it 
gets the timestamp $e.v$ as \\
\[ e.v = max(p.v, q.v); e.v[e.thread]++; \]
Both $p$ and $q$ update their vector to $e.v$.

Object-based vector clocks maintain a vector $V$ of size $|Q|$ with each process 
and object. Whenever, a process $p$ executes an operation $e$ on object $q$ it
gets the timestamp $e.v$ as \\
\[ e.v = max(p.v, q.v); e.v[e.object]++; \]
Both $p$ and $q$ update their vector to $e.v$.

In this paper, we design a vector clock called {\em mixed vector-clock} that uses a combination of 
processes and objects for its components. Clearly, process-based and object-based
vector clocks are special case of our scheme. Moreover, the total number of components
in a mixed vector clock is always less than or equal to the minimum of the
process and object based vector clocks.

\section{An Offline Algorithm}
In this section, we first introduce the thread-object bipartite graph for a computation. Next, we give an offline algorithm to compute the optimal mix-vector-clock by computing the maximum bipartite matching of the thread-object bipartite graph and obtaining a minimum vertex cover. Then we show the mix-vector-clock given by this offline algorithm is optimal in terms of size. 

\subsection{The Thread-object Bipartite Graph}
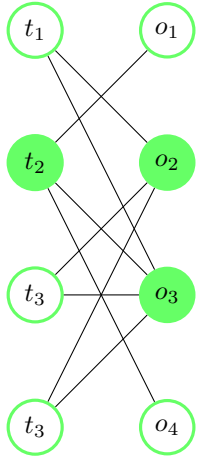
\begin{figure}[htbp] 
\centering
\begin{tikzpicture}[
roundnode/.style={circle, draw=green!60, very thick, minimum size=7mm},
covernode/.style={circle, draw=green!60, fill=green!60, very thick, minimum size=7mm},
]

\node[roundnode]			(t1)       					    {$t_1$};
\node[covernode]			(t2)        [below=of t1] 		{$t_2$};
\node[roundnode]			(t3)			[below=of t2]		{$t_3$};
\node[roundnode]			(t4)			[below=of t3]		{$t_3$};
\node[roundnode]			(o1)			[right=of t1]		{$o_1$};
\node[covernode]			(o2)			[right=of t2]		{$o_2$};
\node[covernode]			(o3)			[right=of t3]		{$o_3$};
\node[roundnode]			(o4)			[right=of t4]		{$o_4$};

\draw[-] (t1) -- (o2);
\draw[-] (t1) -- (o3);
\draw[-] (t2) -- (o1);
\draw[-] (t2) -- (o3);
\draw[-] (t2) -- (o4);
\draw[-] (t3) -- (o2);
\draw[-] (t3) -- (o3);
\draw[-] (t4) -- (o2);
\draw[-] (t4) -- (o3);

\end{tikzpicture} 
\caption{Thread-Object Bipartite Graph of A Computation}
\label{fig:bipar}
\end{figure}

A thread-object computation is composed of events which are in the form of some specific thread doing some operations on a specific object. Notice that such a computation only involves two parties: threads and objects. An operation relates an thread and an object. Therefore, such a computation could be modeled as a bipartite graph if we only focus on the relation between the two parties, i.e, for thread $p$ and object $q$, we only care about whether $p$ has any operation on $q$ or not and ignore exactly how many operations that $p$ has on $q$. If a thread has at least one operation on an object, then there is an edge between them in the bipartite graph. We call such a bipartite graph as thread-object bipartite graph herein. The computation shown in Fig.~\ref{fig:eg} can be converted to the thread-object bipartite graph given in Fig.~\ref{fig:bipar}. The filled vertices represent the minimum vertex cover of this bipartite graph or the components of our \textit{mix-vector-clock}. 

Let $G = \{T \cup O,E\}$ be the thread-object bipartite graph of a given computation. $T$ denote the set of threads involved in the given computation. $O$ denotes the set of objects to which threads in $T$ have operations on in the given computation. For any thread $t \in T$ and object $o \in O$, there is an edge between thread $t$ and object $o$ in the thread-object bipartite graph iff thread $t$ has at least one operation on object $o$ in the given computation.

\subsection{The Offline Algorithm}
Assuming that the thread-object graph representing a computation is given or constructed by the trace generato, we show how to obtain our mix-vector-clock by computing a minimum vertex cover of this bipartite graph. In order to compute such a vertex cover, we use K\"{o}nig-Egerv\'{a}ry Theorem. 
\begin{t1}[K\"{o}nig-Egerv\'{a}ry Theorem]  
In any bipartite graph, the maximum size of a matching equals the minimum size of a vertex cover.
\end{t1}

Based on K\"{o}nig-Egerv\'{a}ry's theorem, we can first compute a maximum matching of the thread-object bipartite graph, which has many algorithms. One simple and efficient such algorithm is the bipartite matching algorithm given by Hopcroft and Karp\cite{HopKar:1973:SIAM}, which achieves a time complexity of $O(n^{5/2})$. 
  
The basics of this matching algorithm is stated as below: at each iteration, this algorithm searches for shortest augmenting paths denoted as $\{Q_1, Q_2,..., Q_t\}$ relative to existing matching $M$ and augment the current matching. The new matching $M'$ is obtained by $M \oplus Q_1 \oplus Q_2 \oplus \cdot\cdot\cdot\oplus Q_t$,  where $\oplus$ represents symmetric difference. When there is no augmenting path in the bipartite graph, the maximum matching is found. The details of this algorithm can be found in\cite{HopKar:1973:SIAM}.

Next, given the maximum matching, we can directly apply K\"{o}nig-Egerv\'{a}ry Theorem to convert the maximum matching to minimum vertex cover so as to get the mix-vector-clock. The pseudocode to compute the mix-vector-clock is given in {\bf Algorithm 1}. 
 
In this algorithm, we first compute a maximum matching $M^*$ of $G$. Given $M*$, the set of unmatched threads can be directly obtained, denote this set as $S$. Line 3-9 is the procedure to convert the maximum matching $M^*$ to a minimum vertex cover. $Z$ is the set of nodes in the graph which are connected by $M^*-$alternating paths to $S$. The minimum vertex cover $C^*$ can be computed as $(T-Z) \cup (O~\cap~Z)$. This procedure can be found in the proof for K\"{o}nig-Egerv\'{a}ry Theorem in \cite{BondyMurty:1976:Book} (Theorem 5.3). For completeness, we include the proof here. 

\begin{algorithm}
\caption{Minimum \textit{Mixed-vector-clock}}
\begin{algorithmic}[1]
\STATE Finding a maximum matching of $G$, denoted as $M*$
\STATE Compute the set of unmatched threads, denoted as $S$
\STATE Let $Z$ := $S$
\FOR{ $s \in S$}
\STATE Start from $s$, BFS search via alternating paths
\STATE Let $B_s$ denote the set of vertexes traversed by BFS 
\STATE $Z$ := $Z \cup B_s$ 
\ENDFOR
\STATE Return $C^{*} = (T-Z) \cup (O~\cap~Z)$
\end{algorithmic}
\end{algorithm}


\begin{l1}
{\bf Algorithm 1} computes a minimum vertex cover of a graph. 
\begin{IEEEproof}
First, let's show that $C^{*}$ is a vertex cover. Let $(t, o)$ be any edge from $E$. There are two possible cases.

Case 1: Edge $(t, o)$ belongs to an alternating path $p \in P$,  then it's right endpoint is in $C$.

Case 2: Edge $(t, o)$ does not belong to any alternating path. If $(t, o)$ is matched, then it's left endpoint $t$ couldn't be in any alternating path (otherwise $(t, o)$ belongs to such an alternating path. Thus, $t \in (T-L)$. If $(t, o)$ is unmatched, then its left endpoint $t$ cannot be in any alternating path, for such a path could be extended by adding $(t, o)$ to it.  

Second, let's prove $C^{*}$ is minimum.
Every vertex in $C^{*}$ is matched. For, every vertex in $(T-L$ is matched because $L$ is a superset of $S$, the set of unmatched left vertices. And every vertex in  $(O~ \cap ~ L)$ must also be matched, for if there existed an alternating path to an unmatched vertex then changing the matching by removing the matched edges from this path and adding the unmatched edges in their place would increase the size of the matching. However, no matched edge can have both of its endpoints in $C^{*}$ . Thus, $C^{*}$ is a vertex cover of cardinality equal to M, and must be a minimum vertex cover.
\end{IEEEproof}
\end{l1}

Given the minimum vertex cover of the thread-object bipartite graph, the mix-vector-clock is simply constructed by assigning each thread or object in the minimum vertex cover as a component in the vector clock.

\subsection{Timestamping Events Using Mix-vector-clock}
The offline algorithm gives a mix-vector-clock. To timestamp events in a thread object computation, we let each thread and each object keep a mix-vector-clock. Now let us look at how each thread modify its mix-vector-clock in order to track causality of operations. For thread $p$, while performing operation $e$ on object $q$, thread $p$ needs to check whether itself or object $q$ is in the mix-vector-clock and increases the component correspondingly. Here is how to update the timestamp of event $e$:
\[ \bm{if~ q \in v~ then:} e.v = max(p.v, q.v); e.v[e.object]++; \]
\[ \bm{if~ p \in v~then:} e.v = max(p.v, q.v); e.v[e.thread]++; \]
Both thread $p$ and object $q$ update their mix-vector-clock to be $e.v$.

\begin{figure}[htbp]
\centerline{\includegraphics[width=0.5\textwidth]{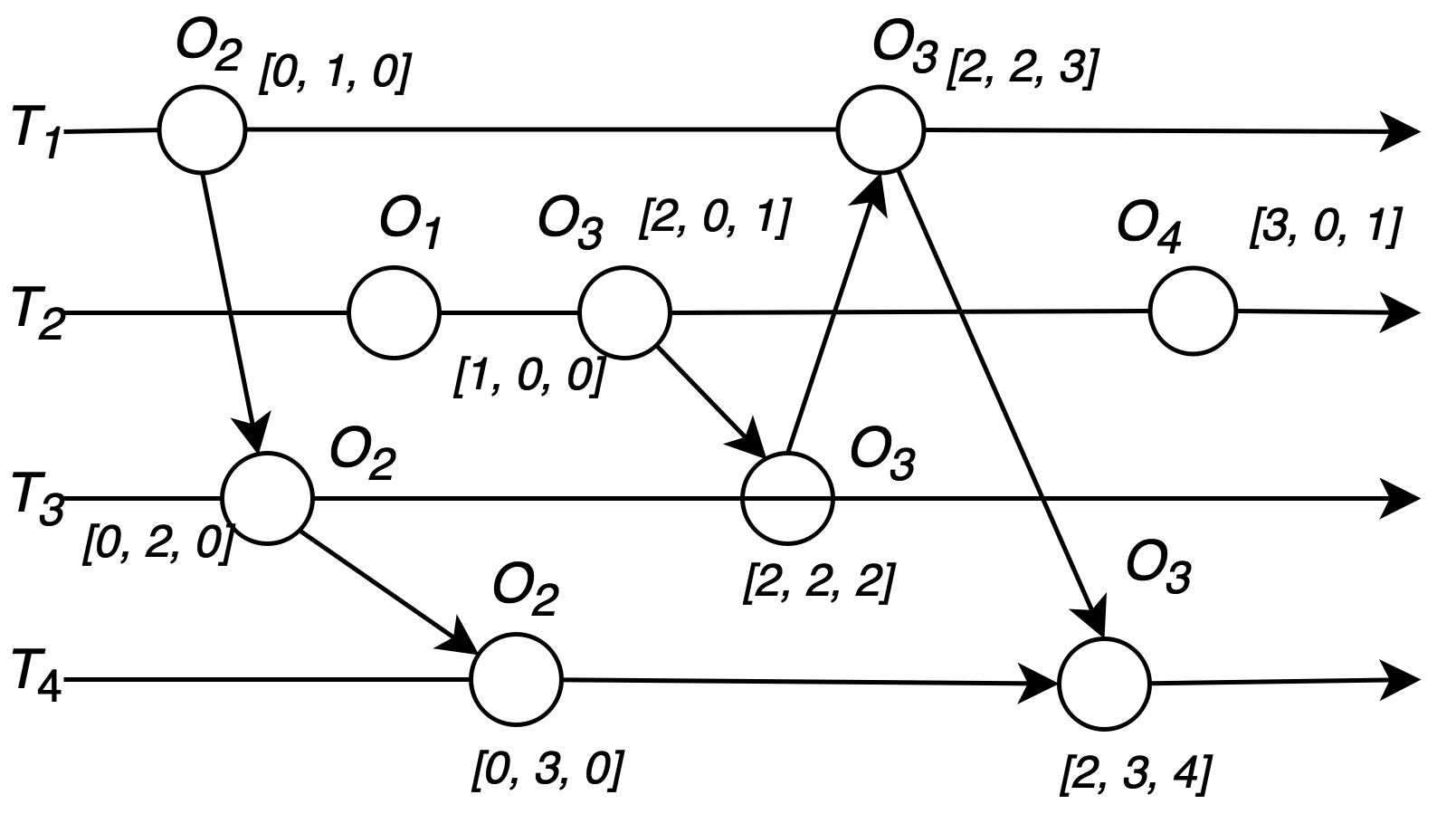}}
\caption{Timstamping Events Using Mix-vector-clock}
\label{fig:timestamp}
\end{figure}

Fig.~\ref{fig:timestamp} shows the timestamp for each event in the computation given in Fig. \ref{fig:eg}.  The components in the mix-vector-clock correspond to thread $T_2$, object $O_2$ and object $O_3$, respectively. Initially, the \textit{mix-vector-clock} for all threads and objects are $[0,~0, ~0]$. Let $[p, q]$ represents the event that thread $p$ performs an operation on object $q$. In Fig. \ref{fig:timestamp}, the mix-vector-clock of $[T_2, O_1]$ is smaller than vector of $[T_3, O_3]$, so we have $[T_2, O_1] \rightarrow [T_3, O_3]$. From the computation, we know that $[T_2, O_1] \rightarrow [T_2, O_3]$ and $[T_2, O_3] \rightarrow [T_3, O_3]$, we also have that $[T_2, O_1] \rightarrow [T_3, O_3]$. 

\subsection{Proof of Correctness and Optimality}
Let $e$ be an operation in the computation, $e.p$ be the associate process of $e, e.q$ be the associate object of $e$, $e.c$ be the component which is in the mix-vector-clock. $e.c = e.p \wedge e.q$. Let $s$ and $t$ be any two operations in the computation. We show the correctness and optimality by the following Lemma and Theorems.
\begin{l1} \label{lem:happen-before}
Let $s \neq t$. Then, 
 \[s \nrightarrow t \Rightarrow t.v[s.c] < s.v[s.c] \]
 
\begin{IEEEproof}
If $s.p = t.p$, then it follows that $t \prec s$. Before executing operation $s$, the vector clock kept in process $p$ must be no less than $t.v$, since $p$ updates its vector to be the same as $t.v$ after operation $t$. At operation $s$, $s.v = max(s.p.v, s.q.v) and s.v[s.c]++$, thus $t.v[s.c] < s.v[s.c]$. Otherwise, $s.p \neq t.p$. Since $s$ will increase $s.v[s.c]$ will increase its component by at least one and this increase could not have been seen by $t$, since $s \nrightarrow t$, it follows that $t.v[s.c] < s.v[s.c]$. 
\end{IEEEproof}
\end{l1}

\begin{t1} \label{th:correctness}
(\textit{Correctness:}) The \textit{mix-vector-clock} is a valid vector clock.

\begin{IEEEproof}
In order to show the \textit{mix-vector-clock} is a valid vector clock,  we need to show that it satisfies the constraint: $ \forall s,t: s \rightarrow t \Leftrightarrow s.v < t.v$.

($\Rightarrow$) $s \rightarrow t \Rightarrow s.v < t.v$

Case 1: $s.p = t.p$ 
As we can see from the vector clock algorithm, within the same thread, each event will first set its new vector clock to be the max of previous vector clock and vector on object. Thus $s.v < t.v$.
Case 2: $s.q = t.q$. $s$ immediately precedes $t$ in the sequence of events on object $s.q$. Therefore, $s.v[s.q] = t.v[s.q] + 1$ and for $\forall j \neq s.q$: $s.v[j] = t.v[j]$. Thus, $s.v < t.v$.

($\Leftarrow$) $s \nrightarrow t \Rightarrow s.v \nless t.v$

From Lemma \ref{lem:happen-before}, we have $s \nrightarrow t \Rightarrow t.v[s.c] < s.v[s.c] $. Thus, $s.v \nless t.v$.
\end{IEEEproof}
\end{t1}

\begin{t1} \label{th:optimality}
(\textit{Optimality:}) The \textit{mix-vector-clock} is optimal in size.

\begin{IEEEproof}
The offline algorithm makes use of the components of minimum vertex cover $v$ of the thread-object bipartite graph as the \textit{mix-vector-clock}, which is also a valid vector clock of the computation. We need to show this $v$ is minimum in size. Since timestamping all events in a computation requires the vector clock being able to order each event. Suppose there exists a smaller vector clock $v'$ which timestamps all events. The fact that events in computation corresponds to edges in its bipartite graph  indicates that $v'$ is also a vertex cover of the bipartite graph, thus contradicts $|v| \leq |v'|$. So, the \textit{mix-vector-clock} obtained by the offline algorithm is optimal in size.  
\end{IEEEproof}
\end{t1}

\section{Mix-vector-clock for Online Computation}
In this section, we consider the case when the computation is not given in advance; instead, each event of the computation is revealed in an online fashion. We assume that only one event is revealed at any time. Thus, in the online setting, we need to maintain a valid dynamic vector clock when events of a computation arrive one at a time. The thread-object bipartite graph for the computation changes when events arrive in the online fashion. When an event $e = (t, o)$ is revealed, there could be two cases. Either the thread $t$ has already performed some operation on object $o$, i.e, there is already an edge between $t$ and $o$ in the current thread-object bipartite graph. In this case, the thread-object bipartite graph does not change. For the other case, thread $t$ has never performed any operation on object $o$, i.e, there is no edge between $t$ and $o$ in the current thread-object bipartite graph. In this case, an edge between $t$ and $o$ should be added into the thread-object bipartite graph. Note that in the online setting the idea of using minimum vertex cover to be the components of mix-vector-clock cannot be applied, since the minimum vertex cover of the thread-object graph is changing and the existing components in a mix-vector-clock should not be modified as a new event arrives.
That is, we can only add new components to the vector clock.

The conventional naive solution is to always choose thread or always choose object as components of the vector clock as a new event occurs. This mechanism would result in a vector clock with size equal to the number of threads or objects for all computations. Another intuitive mechanism is to randomly choose the object or the thread to add into the vector clock with equal probability. Notice that the hardness of timestamping an online computation stems from the unpredictability of future events. We can only estimate the future using information we already have. Therefore, we propose another mechanism which makes use of the partial computation occurred so far to predict the future events.  Specifically, when a new event occurs, if the associated object is more "\textit{popular}" than the associated thread, then we choose the object, otherwise, we choose the thread. We propose the definition of popularity as below. 

\begin{d1}
The \textit{popularity} of a vertex $v$ in a bipartite graph $G = \{U \cup V,E\}$ is
$pop(v) = \frac{deg_v}{|E|} $, where $deg_v$ is the degree of vertex $v$ and $|E|$ is the total number of edges in the graph. We say one node is more popular than another node if it has higher \textit{popularity}.
\end{d1} 

We assume that only one event can occur at a single time. An event comes in with its associated thread and object. We are not supposed to modify components existing in the mix-vector-clock. The three mechanism are formally listed as:

1. \textit{Naive:} Always choose threads or objects.

2. \textit{Random:} Randomly choose the associated object or thread of the new event with equal probability.

3. \textit{Popularity:} Based on popularity of threads and objects. When a new event comes in. If one of the associated thread or object is already in the vector clock, the the vector clock remains same. Otherwise, compute the popularity of the associated thread and object, add the one with higher popularity into the vector clock.  

\section{Evaluation}
In this section, to evaluate the performance of our offline algorithm and compare the performance of the three mechanisms we proposed for online setting, we consider the following two Scenarios: 

\textit{Uniform}: Evaluating on a uniformly and randomly generated thread-object bipartite graph, i.e, each thread and object have same popularity.

\textit{Nonuniform}: Evaluating on a thread-object bipartite graph in which a small fraction of objects and threads are much more popular than other threads and objects.

The bipartite graph in \textit{Uniform} scenario could be generated by adding a edge between each thread and each object with a same specific probability. For \textit{Nonuniform} scenario, the bipartite graph can be generated by adding an edge between popular threads and objects with a higher probability and non-popular threads and objects with a smaller probability. 

\begin{figure}[h] 
\centerline{\includegraphics[width=0.5\textwidth]{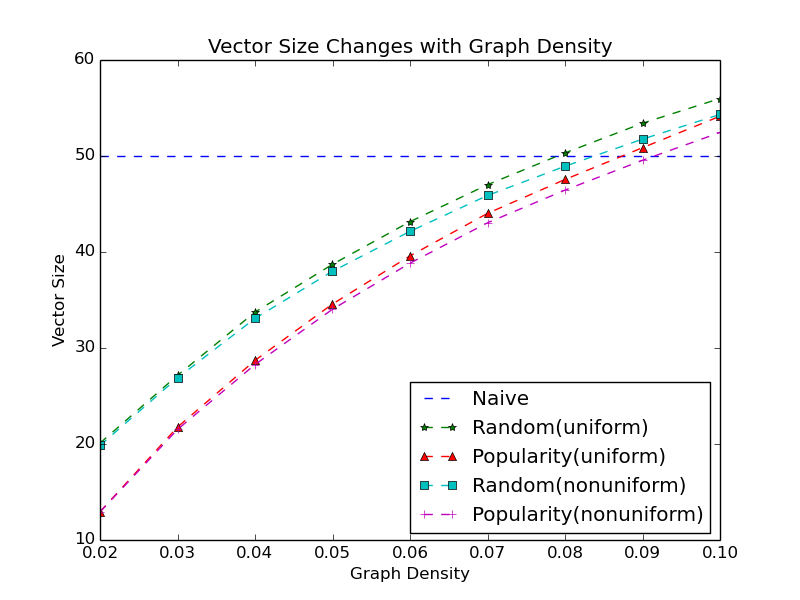}}
\caption{Vector Size Varies as Graph Density Increases.}
\label{fig:offDensity}
\end{figure}

In our first evaluation we consider how graph density affects the vector clock size of the three mechanisms. We set the number of threads and objects in the computation to be 50, respectively, i.e, each side of the thread-object bipartite graph has 50 nodes. We compute the final vector clock size by applying the above three mechanisms to the above two different scenarios, as the density of thread-object bipartite graph increases. The results are shown in Fig. \ref{fig:offDensity}. The first important conclusion we get is that when the density of the bipartite graph is small \textit{Random} and \textit{Popularity} method produce significantly smaller vector clock size than the \textit{Naive} method. However, when the density of graph exceeds a certain threshold, their performance becomes worse than \textit{Naive}. In addition, we found that \textit{Random} and \textit{Popularity} mechanism can obtain much better solution in the \textit{Nonuniform} case than the \textit{uniform} case. Thus, we can conjecture that those two methods are better suited in the computation in which some objects or threads are more popular than other objects and threads. Comparing performance of \textit{Random} and \textit{Popularity}, we found \textit{Popularity} is a slightly better than \textit{Random}. This can be explained by the fact that by choosing popular nodes as vector clock component, we can cover more edges. Thus, the vector clock size would be smaller.

\begin{figure}[h] 
\centerline{\includegraphics[width=0.5\textwidth]{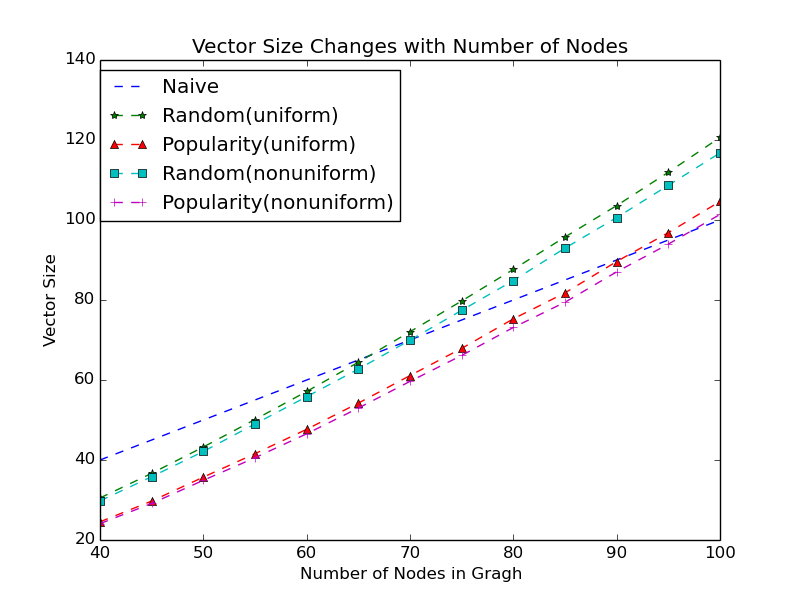}}
\caption{Vector Size Varies as Number of Nodes Increases}
\label{fig:offNodes}
\end{figure}

In our second evaluation, we fix the density of graph to be 0.05 and evaluate the performance of the three mechanisms as we increase the number of nodes in the thread-object bipartite graph. From Fig. \ref{fig:offNodes}, we can notice that as the number of nodes in the graph increases, i.e,  the number of threads and objects increases in a computation, the vector size increases. When the number of nodes is below a certain threshold, 70 here, \textit{Random} and \textit{Popularity} generates smaller vector clock size than \textit{Naive}. Once the number of nodes exceeds that threshold, \textit{Naive} is better, which means by simply choosing either all threads or objects as vector clock components gives a smaller vector clock. Therefore, we can get the conclusion that these two techniques are more effective in simple computations, i.e, in computations which involves relatively small number of threads and objects.

\begin{figure}[h] 
\centerline{\includegraphics[width=0.5\textwidth]{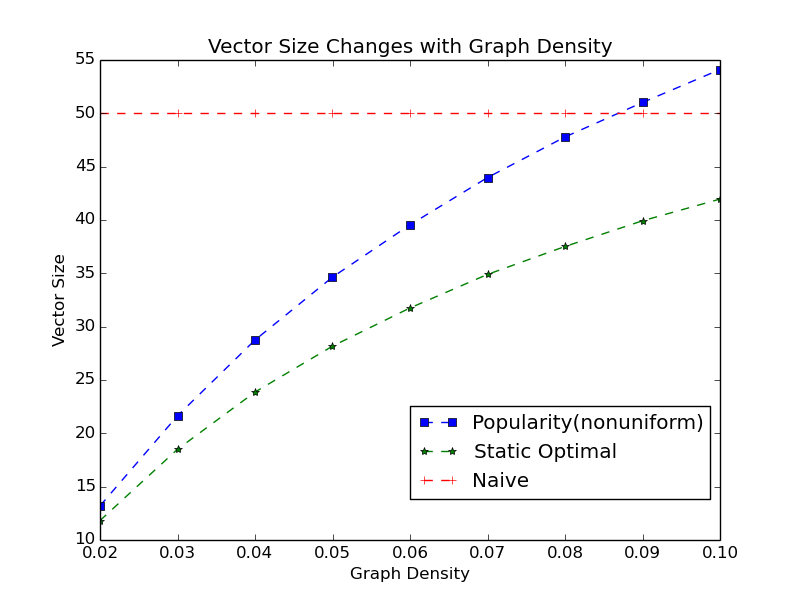}}
\caption{Vector Size Varies as Graph Density Increases}
\label{fig:staticDensity}
\end{figure}

\begin{figure}[h] 
\centerline{\includegraphics[width=0.5\textwidth]{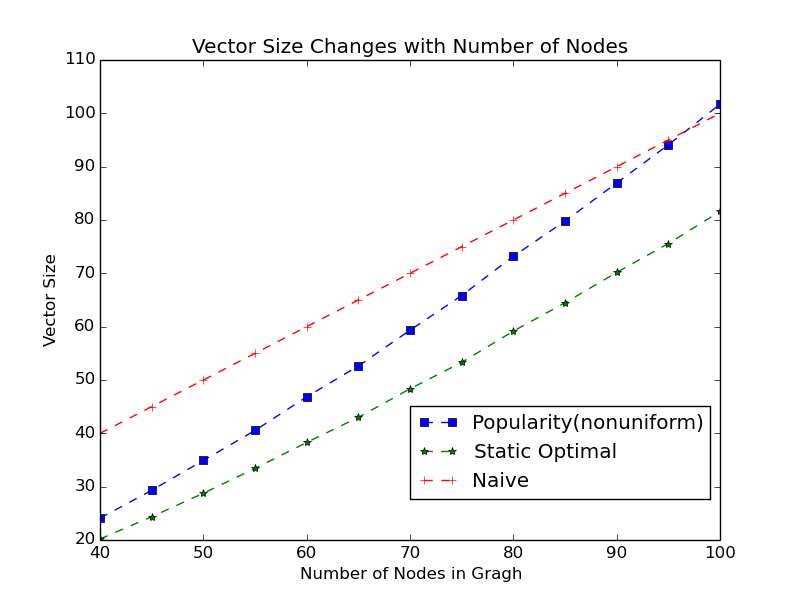}}
\caption{Vector Size Varies as Number of Nodes Increases}
\label{fig:staticNode}
\end{figure}

In our third evaluation, we want to know how far the online case drifts from the static case. We choose the \textit{Popularity} mechanism for the online case. We also consider the \textit{Naive} mechanism , which can both be applied to online case and static case and generates the same vector clock. For the static case, we use our offline algorithm proposed in Section III. For the experiment, we first apply the \textit{Popularity} mechanism as we reveal the edge of the graph one by one. Then, after we have the whole graph, we apply the offline algorithm. Fig. \ref{fig:staticDensity} shows the results we get when we set the number of nodes to be 50 and increase graph density. Fig. \ref{fig:staticNode} shows the results we get when we fix the density to be 0.05 and increase the number of nodes in the graph. First, we can notice that our offline algorithm generates a vector clock with significantly smaller size than the \textit{Naive} solution. For example, \textit{Naive} has a vector clock of size 50 and the offline algorithm reduces that to be around 35, when there are 50 threads and the graph density is 0.05. Besides, in online setting, although the \textit{Popularity} mechanism cannot completely achieve as small vector size as the optimal solution does, the gap is within a reasonable number. For example, \textit{Popularity} generates a vector of size round 56 while the optimal is round 48, when there are 70 threads and the graph density is 0.05. Also, as graph density or number of nodes in graph increases, the gap is increasing which again indicates the \textit{Popularity} mechanism is not suitable for relatively dense graph. 

Combining the results of the above evaluations, for the online problem, a practical mechanism would be to set thresholds for both graph density and number of nodes in graph. At the beginning, we adopt the \textit{Popularity} mechanism and as more events come in we adopt the \textit{Naive} approach if the graph parameters exceeds the thresholds.   
 
\section{Related Work}

Several techniques have been proposed to reduce the overhead imposed
by Fidge/Mattern's vector clocks
\cite{Fid:1988:ACSC,Fid:1989:WPDD,Fid:1991:Com,Mat:1989:WDAG}. Singhal
and Kshemkalyani \cite{SinKsh:1992:IPL} present a technique to reduce
the amount of data piggybacked on each message. The main idea is to
only send those entries of the vector along with a message that have
changed since a message was last sent to that process.
 H{\'e}lary
\emph{et al} \cite{HelRay+:2003:KDE} further improve upon Singhal and
Kshemkalyani technique and describe a suite of algorithms that
provide different trade offs between space overhead and communication
overhead. The ideas described in the two papers are actually
orthogonal to the ideas presented in this paper and, therefore, can
also benefit our timestamping algorithm by reducing its overhead.

Agarwal and Garg \cite{AgaGar:2005:PODC} have proposed a
class of logical clock algorithms, called chain clock,
for tracking dependencies between {\em relevant} events based on generalizing a
process to any chain in the computation poset. Their algorithm reduces
the number of components required in the vector when the set of relevant
events is a small fraction of the total events. Our work is mot closely related to this work.
They provide two different algorithms: the first algorithm adds any newly arrived event to a chain 
with the guarantee that no more that $|P|$ chains are necessary. The second algorithm
uses online chain decomposition of a poset to guarantee that no more that 
$(w+1)w/2$ chains are necessary where $w$ is the width of the poset.
In this paper, our algorithm uses components that are either for the process or for the object
and guarantees that the number of components is never more than $min(|P|, |Q|)$.

Garg, Skawratananond, and Mittal \cite{DBLP:journals/dc/GargSM07} have proposed an algorithm
to timestamp messages in a distributed system. They assume that all messages are synchronous and show that
such systems can have timestamps of vector clocks with dimension less than $N$. They define the notion of
an undirected communication graph with the set of vertices as processes and the edges denoting which processes
can communicate. They show that the number of components required is equal to the number of stars and triangles
the communication graph can be decomposed into. 
Our technique of using
vertex cover is inspired from that work even though their work is strictly for distributed systems and they 
do not consider mixed-clocks.
We have two types of entities in our system ---
threads and objects and the dimension of the vector clock reduces to a vertex cover of the bipartite graph
that represents the interaction between these entities. 

\section{Conclusions}
This paper proposes an offline algorithm to compute a mixed vector clock composed of a mix of threads and objects, which is shown to be a correct vector clock and have optimal size, to timestamp events in a computation. Thread-object bipartite graph is constructed based on the given computation and then the minimum vertex cover of this bipartite graph is computed and the threads and objects in this vertex cover are adopted as the components of the mix vector clock. In a online computation in which events are coming one by one, two mechanisms are proposed, the \textit{Random} mechanism which randomly choose the associated thread or object as vector clock component and the \textit{Popularity} mechanism which choose the thread or object based on their popularity. By evaluating on thread-object bipartite graphs with different characteristics, we get the conclusion that the \textit{Popularity} mechanism shows best performance on \textit{non-uniform} graphs. 

\bibliography{citations}

\end{document}